# *Spontaneous Parity-Time Symmetry Breaking in Moving Media*

*Mário G. Silveirinha*[*]

*(1)University of Coimbra, Department of Electrical Engineering – Instituto de Telecomunicações, Portugal, mario.silveirinha@co.it.pt*

**Abstract**

Optical instabilities in moving media are linked to a spontaneous parity-time symmetry breaking of the system. It is shown that in general the time evolution of the electromagnetic waves in moving media is determined by a non-Hermitian parity-time symmetric operator. For lossless systems the frequency spectrum of the time evolution operator may be complex valued, and has a mirror symmetry with respect to the real-frequency axis. The possibility of optical amplification of a light pulse in the broken parity-time symmetry regime is demonstrated.

---

[*] To whom correspondence should be addressed: E-mail: *mario.silveirinha@co.it.pt*

# I. Introduction

Unbroken parity time ($\mathcal{PT}$) - symmetric Hamiltonians have been suggested as the basis for a new class of generalized complex quantum theories that do not require the Hamiltonian to be Hermitian [1-2]. The discovery that $\mathcal{PT}$ - symmetric Hamiltonians can be used to describe the physical reality – without violating the condition that the time evolution is unitary and the reality of the energy eigenvalues – contributed to deepen the understanding of the quantum mechanics foundations and extended the set of acceptable theories to cases previously judged as unphysical. A recent article discussed some possible limitations of $\mathcal{PT}$ - symmetric theories as fundamental theories of nature [3].

$\mathcal{PT}$ -symmetric Hamiltonians are often defined parametrically, such that a specific physical theory is associated with a parameter that measures the departure from the Hermiticity condition. Notably, $\mathcal{PT}$ - symmetric systems can undergo an abrupt phase transition, such that beyond a critical parameter threshold a spontaneous symmetry breaking occurs and the energy spectrum becomes complex-valued. Most remarkable, in such conditions the eigenfunctions of the Hamiltonian ($\hat{H}$) do not have to be eigenfunctions of the $\mathcal{PT}$ -operator despite the fact the operators commute [1-2]. The reason is that the $\mathcal{PT}$ operator is antilinear, and thus $\left[\hat{H}, \mathcal{PT}\right] = 0$ does not imply that the operators are simultaneously diagonalizable.

$\mathcal{PT}$ - symmetric Hamiltonians have also been previously studied and realized at optics (in the framework of classical physics) through a judicious inclusion of gain/loss regions [4-7]. It has been shown that such systems can exhibit power oscillations, double refractions, non-reciprocal wave propagation, and are characterized by phase transitions beyond which they can become unstable. A transformation optics design of $\mathcal{PT}$-symmetric photonic structures was reported in Ref. [8]. The application of $\mathcal{PT}$-symmetric methods to metamaterials was also discussed in other works [9, 10, 11].

In a different line of research, we have recently shown [12, 13, 14] that moving media may support wave instabilities, such that if the relative velocity of the bodies exceeds a certain threshold the system may become unstable and may start spontaneously emitting light. It was shown that these wave instabilities in *noncharged* polarizable moving media are strictly linked to the Cherenkov and Smith-Purcell effects [15, 16]. The wave dynamics in these systems has several peculiarities. First, provided the speed of the moving bodies is enforced to remain time independent, the system may support exponentially growing oscillations, even in presence of realistic material loss and dispersion [12, 13, 14]. Second, notwithstanding the exponential growth of the wave fields, in the absence of external electromagnetic sources the wave energy and momentum are conserved. Furthermore, the wave energy density has no lower bound and can be negative. These seemingly absurd and counterintuitive properties were discussed in detail in Refs. [12, 13, 14]. It was proven that they do not contradict any physical laws, and that when the mechanical degrees of freedom of the system are properly taken into account the total energy density of the system is always greater than zero. Crucially, it was demonstrated that a friction-type force acts on the moving bodies to oppose their relative motion. Thus, the wave instabilities result from the conversion of kinetic energy into electromagnetic energy. Plasma wave instabilities due to the drift of electrons in semiconductors have been reported in other works [17, 18]. Also, electron-scale instabilities where the shear kinetic energy flow is converted into electric and magnetic field energy have been studied in Refs. [19, 20, 21]. It has been suggested that these shear instabilities may be an important dissipation mechanism in astrophysical jets [21].

Furthermore, in Refs. [12, 13, 22] we proved that the Cherenkov-type electromagnetic instabilities in moving media are strictly linked to noncontact quantum friction, and a fully quantum mechanical theory for this effect at zero-temperature was developed. A related quantum Cherenkov friction effect was discussed in Refs. [23-24]. Generically, quantum friction predicts that two perfectly smooth noncharged moving surfaces separated by a vacuum can experience a force of quantum origin that tends to counteract the relative motion [25-33]. Quantum friction can also occur in other scenarios, involving for example rotating dielectric bodies [34-37]. Quantum radiation by moving mirrors [38, 39] and its connections with sonoluminesce have also been extensively studied [40-42].

In this work, we link our previous studies of wave instabilities in moving media with $\mathcal{PT}$ - symmetry methods, and demonstrate that beyond a certain velocity threshold there is a phase transition and a spontaneous parity-time symmetry breaking of the system. We characterize in detail the electromagnetic fields and the wave momentum density associated with the natural oscillations of the system in the broken $\mathcal{PT}$ - regime. Finally, we numerically study the time-domain evolution of the electromagnetic fields emitted by a line source in presence of wave instabilities, demonstrating that the emitted fields are amplified in the broken $\mathcal{PT}$ - regime. The theory of this work is based on classical electrodynamics.

## II. Wave dynamics in moving media

Here, we are interested in the wave interactions in a system formed by two moving slabs (infinitely extended along the $x$ and $y$ directions) and separated by a vacuum region with thickness $d$ (Fig. 1). The two bodies move with speed $\mathbf{v}_i = v_i \hat{\mathbf{x}}$ ($i=1,2$) with respect to some inertial reference frame (laboratory frame). We suppose that the speeds $v_i$ are enforced to be time independent. As discussed in detail in our previous works [12, 13], having $dv_i/dt = 0$ may require that either *(i)* an external mechanical force is applied to the bodies to counteract radiation induced forces or *(ii)* the mass density of the bodies is very large. Notably, when $dv_i/dt = 0$ the electrodynamics of the system becomes uncoupled from the equations of motion associated with the mechanical degrees of freedom (see Appendix A of Ref. [13]), and hence a fully relativistic treatment of the problem is possible. In particular, for lossless dispersionless media the relativistic relation between the classical $\mathbf{D}$ and $\mathbf{B}$ fields and the classical $\mathbf{E}$ and $\mathbf{H}$ fields is [43, 44]:

$$\begin{pmatrix} \mathbf{D} \\ \mathbf{B} \end{pmatrix} = \begin{pmatrix} \varepsilon_0 \bar{\bar{\varepsilon}} & \frac{1}{c}\bar{\bar{\vartheta}} \\ \frac{1}{c}\bar{\bar{\zeta}} & \mu_0 \bar{\bar{\mu}} \end{pmatrix} \begin{pmatrix} \mathbf{E} \\ \mathbf{H} \end{pmatrix} \equiv \mathbf{M} \cdot \begin{pmatrix} \mathbf{E} \\ \mathbf{H} \end{pmatrix} \tag{1}$$

where the dimensionless parameters $\bar{\bar{\varepsilon}}$, $\bar{\bar{\mu}}$, $\bar{\bar{\vartheta}}$ and $\bar{\bar{\zeta}}$ are such that,

$$\bar{\bar{\varepsilon}} = \varepsilon_t \left( \bar{\bar{\mathbf{I}}} - \hat{\mathbf{x}}\hat{\mathbf{x}} \right) + \varepsilon \hat{\mathbf{x}}\hat{\mathbf{x}}, \qquad \varepsilon_t = \varepsilon \frac{1-\beta^2}{1-n^2\beta^2} \tag{2a}$$

$$\bar{\bar{\mu}} = \mu_t \left( \bar{\bar{\mathbf{I}}} - \hat{\mathbf{x}}\hat{\mathbf{x}} \right) + \mu \hat{\mathbf{x}}\hat{\mathbf{x}}, \qquad \mu_t = \mu \frac{1-\beta^2}{1-n^2\beta^2} \tag{2b}$$

$$\bar{\bar{\zeta}} = -\bar{\bar{\vartheta}} = -a\hat{\mathbf{x}} \times \bar{\bar{\mathbf{I}}}, \qquad a = \beta \frac{n^2-1}{1-n^2\beta^2} \tag{2c}$$

where $\beta = v/c$, $n^2 = \varepsilon\mu$, and $\varepsilon$ and $\mu$ are the material parameters in the rest frame co-moving with the pertinent body. It will be shown in Sect. VI that Eq. (1) can be generalized to dispersive media. The dynamics of the electromagnetic field is determined by the Maxwell's equations which can be written in a compact form as:

$$\hat{N} \cdot \mathbf{F} = i\mathbf{M} \cdot \frac{\partial \mathbf{F}}{\partial t} + i\mathbf{j}_{ext}, \quad \text{with } \hat{N} = \begin{pmatrix} 0 & i\nabla \times \\ -i\nabla \times & 0 \end{pmatrix}. \quad (3)$$

where $\mathbf{F} = (\mathbf{E} \ \mathbf{H})^T$, $\mathbf{G} = (\mathbf{D} \ \mathbf{B})^T$, $T$ denotes the matrix (or vector) transpose, and $\mathbf{M} = \mathbf{M}(z)$ is the material matrix defined by Eq. (1). The six-component vector $\mathbf{j}_{ext} = (\mathbf{j}_{e,ext} \ 0)^T$ is written in terms of a hypothetical external electric current density ($\mathbf{j}_{e,ext}$). For future reference, we note that the instantaneous *wave* energy density is:

$$W_{EM,P} = \frac{1}{2}(\mathbf{E}^* \cdot \mathbf{D} + \mathbf{H}^* \cdot \mathbf{B}) = \frac{1}{2}\mathbf{F}^* \cdot \mathbf{M} \cdot \mathbf{F}. \quad (4)$$

The symbol "*" denotes complex conjugation, and can be ignored for real-valued fields. To fully explore the analogy with quantum physics where the wave function is complex-valued, we allow the electromagnetic fields to be complex-valued.

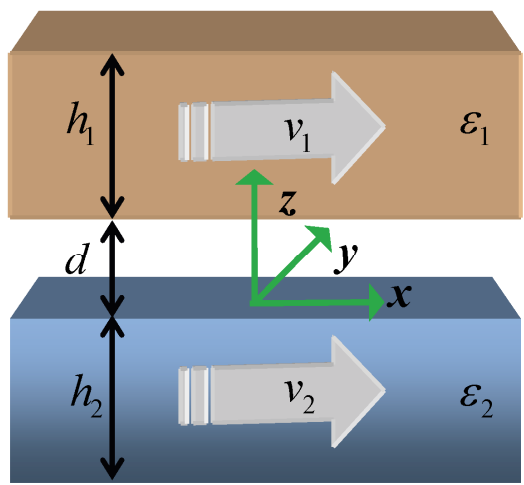

Fig. 1. (Color online) Geometry of the system: two infinitely extended (in the *x* and *y* directions) material slabs separated by an air gap with thickness *d* are in relative motion.

We also introduce the instantaneous electromagnetic and wave momentum densities defined by, respectively,

$$\mathbf{g}_{EM} = \frac{1}{c^2}\frac{1}{2}\left(\mathbf{E}\times\mathbf{H}^* + \mathbf{E}^*\times\mathbf{H}\right), \tag{5a}$$

$$\mathbf{g}_{wv} = \frac{1}{2}\left(\mathbf{D}^*\times\mathbf{B} + \mathbf{D}\times\mathbf{B}^*\right). \tag{5b}$$

Again, for real-valued fields the complex conjugation symbol is irrelevant. The electromagnetic (Abraham) momentum can be regarded as the light momentum. On the other hand, the wave (Minkowski) momentum has both matter and light components [12, 13, 45, 46]. As discussed by Barnett [45], the total momentum of a material medium can be decomposed as $\mathbf{g}_{tot} = \mathbf{g}_{kin} + \mathbf{g}_{EM} = \mathbf{g}_{can} + \mathbf{g}_{wv}$, where $\mathbf{g}_{kin}$ and $\mathbf{g}_{can}$ are the kinetic and canonical momenta densities of the medium matter [47], respectively.

The wave flow in the material media originates a stress along the *x*-direction, such that for the *i*-th body (assumed to be invariant to translations along the *x*-direction) [13]:

$$F_i^{mat} = \frac{dp_{ps,i}}{dt}. \tag{6}$$

In the above, $p_{ps,i} = p_{wv,i} - p_{EM,i}$ is the *x*-component of the so-called pseudo-momentum of the *i*-th slab, and $p_{wv,i}$ and $p_{EM,i}$ are *x*-components of the wave and electromagnetic momenta:

$$\mathbf{p}_{wv,i} = \int_{V_i}\mathbf{g}_{wv}\,d^3\mathbf{r}, \qquad \mathbf{p}_{EM,i} = \int_{V_i}\mathbf{g}_{EM}\,d^3\mathbf{r}, \tag{7}$$

and $V_i$ is the volume of the *i*-th body. In our problem, the stress $F_i^{mat}$ acts to oppose the relative motion of the two slabs and hence it is a friction-type force. The external force required to counterbalance the stress caused by the wave flow is given by $F_{i,x}^{ext} = -F_i^{mat} = -dp_{ps,i}/dt$ [13].

## III. Parity-time symmetry

To establish a precise link between our classical framework, and the $\mathcal{PT}$ - symmetric theories of quantum physics, we introduce a time reversal operator $\mathcal{T}$

$$\mathbf{F}(\mathbf{r}) \xrightarrow{\mathcal{T}} \begin{pmatrix} \mathbf{1} & 0 \\ 0 & -\mathbf{1} \end{pmatrix} \cdot \mathbf{F}^*(\mathbf{r}), \tag{8}$$

and a parity transformation $\mathcal{P}$

$$\mathbf{F}(\mathbf{r}) \xrightarrow{\mathcal{P}} \begin{pmatrix} \mathbf{R}_z & 0 \\ 0 & \mathbf{R}_z \end{pmatrix} \cdot \mathbf{F}(\mathbf{R}_z \cdot \mathbf{r}). \tag{9}$$

where $\mathbf{F} = \begin{pmatrix} \mathbf{E} & \mathbf{H} \end{pmatrix}^T$ and $\mathbf{R}_z = -(\hat{\mathbf{x}}\hat{\mathbf{x}} + \hat{\mathbf{y}}\hat{\mathbf{y}}) + \hat{\mathbf{z}}\hat{\mathbf{z}}$ is the transformation matrix associated with a two-fold rotation about the $z$-axis. Note that the coordinates are transformed as $(x, y, z) \to (-x, -y, z)$ under the considered parity transformation, and thus only the $x$- and $y$-coordinates flip sign. Often the parity transformation is understood as a transformation that flips the sign of all coordinates, different from what is considered here. It is important to note that the $\mathcal{T}$-operator is not linear due to the complex conjugation operation. Moreover, the $\mathcal{T}$ and the $\mathcal{P}$ operators are defined so that the electromagnetic fields are transformed consistently with the usual rules of time-reversal and parity transformations of classical electrodynamics [48].

In the absence of an external source ($\mathbf{j}_{ext} = 0$), the equation that describes the dynamics of the wave fields (3) can be rewritten in a form alike to the Schrödinger equation with $\hbar = 1$:

$$\hat{H}_{cl} \cdot \mathbf{F} = i \frac{\partial \mathbf{F}}{\partial t}, \qquad \text{where} \quad \hat{H}_{cl} = \mathbf{M}^{-1} \cdot \hat{N}. \tag{10}$$

Thus, the operator $\hat{H}_{cl}$ describes the time evolution of the classical fields, analogous to the Hamiltonian operator in quantum physics. It can be checked that $\hat{H}_{cl}$ has neither the time-reversal symmetry nor the parity symmetry because such symmetries imply flipping the velocities of the moving slabs. In particular, as a consequence of the lack of time-reversal invariance, the electromagnetic response of the moving bodies does not satisfy the Lorentz reciprocity theorem [43].

On the other hand, the $\mathcal{PT}$ symmetry requires flipping the velocities of the slabs twice, and hence our system is expected to be $\mathcal{PT}$ symmetric. This can be explicitly checked by noting that the composition of time-reversal operator $\mathcal{T}$ and of the parity operator $\mathcal{P}$ acts on the electromagnetic fields as

$$\mathbf{F}(\mathbf{r}) \stackrel{\mathcal{PT}}{\rightarrow} \tilde{\mathbf{F}}(\mathbf{r}) \equiv \begin{pmatrix} \mathbf{R}_z & 0 \\ 0 & -\mathbf{R}_z \end{pmatrix} \cdot \mathbf{F}^*(\mathbf{R}_z \cdot \mathbf{r}). \tag{11}$$

Straightforward calculations show that:

$$\hat{N} \cdot \mathcal{PT} = \mathcal{PT} \cdot \hat{N}, \tag{12a}$$

$$\mathbf{M} \cdot \mathcal{PT} = \mathcal{PT} \cdot \mathbf{M}. \tag{12b}$$

where $\hat{N}$ and $\mathbf{M}$ are operators defined as in Eqs. (1)-(3). We used the fact that the system is invariant to translations along the $x$ and $y$–directions, and that $\mathbf{M}$ is real-valued because the dielectrics are lossless. Hence, from the commutation relations (12) it is obvious that $\hat{H}_{cl}$ also commutes with the $\mathcal{PT}$ operator:

$$\left[ \hat{H}_{cl}, \mathcal{PT} \right] \equiv \hat{H}_{cl} \cdot \mathcal{PT} - \mathcal{PT} \cdot \hat{H}_{cl} = 0. \tag{13}$$

Hence, $\hat{H}_{cl}$ is indeed a $\mathcal{PT}$-symmetric operator, and the mapping $\mathbf{F}(\mathbf{r}) \stackrel{\mathcal{PT}}{\rightarrow} \tilde{\mathbf{F}}(\mathbf{r})$ transforms solutions of Maxwell's equations into solutions of Maxwell's equations such that if $\mathbf{F}(\mathbf{r},t)$ is a solution then $\mathcal{PT}\cdot\mathbf{F}(\mathbf{r},-t)$ also is. Interestingly, the transformation $\mathbf{F}(\mathbf{r})\rightarrow\tilde{\mathbf{F}}(\mathbf{r})$ plays a crucial role in the quantization theory developed in our previous work [12], but its connection with the $\mathcal{PT}$ operator was unnoticed. Because the $\mathcal{PT}$-time operator is idempotent one has $\mathbf{F}(\mathbf{r})=\tilde{\tilde{\mathbf{F}}}(\mathbf{r})$.

To further develop the analogy of our system with $\mathcal{PT}$-symmetric Hamiltonians, let us consider a natural mode of oscillation $\mathbf{F}_\omega$ associated with the frequency $\omega$ (time variation is $e^{-i\omega t}$), such that

$$\hat{H}_{cl}\cdot\mathbf{F}_\omega=\omega\mathbf{F}_\omega. \tag{14}$$

Because $\left[\hat{H}_{cl},\mathcal{PT}\right]=0$ we can state that:

$$\hat{H}_{cl}\cdot\tilde{\mathbf{F}}_\omega=\omega^*\tilde{\mathbf{F}}_\omega. \tag{15}$$

Therefore, the $\mathcal{PT}$-transformed eigenfunction is associated with the complex conjugated oscillation frequency. Note that for a transverse to $z$ spatial variation $e^{ik_x x}e^{ik_y y}$ a real-valued transverse wave vector $(k_x,k_y)$ remains invariant under a $\mathcal{PT}$-transformation, and thus if $\mathbf{F}_\omega \leftrightarrow (\omega,k_x,k_y)$ then $\tilde{\mathbf{F}}_\omega \leftrightarrow (\omega^*,k_x,k_y)$.

Consistent with the ideas of quantum physics [1-2], we say that $\hat{H}_{cl}$ has an *unbroken* $\mathcal{PT}$ symmetry if the spectrum of $\hat{H}_{cl}$ is real-valued. In such a case, it is seen from (14)-(15) and from the fact that the $\mathcal{PT}$-operator is idempotent and antilinear, that the eigenfunctions of $\hat{H}_{cl}$ may be chosen to satisfy $\tilde{\mathbf{F}}_\omega=\mathbf{F}_\omega$, i.e. the eigenfunctions can be chosen so that they are invariant under a $\mathcal{PT}$ - transformation.

On the other hand, in case of a *broken* $\mathcal{PT}$-symmetry the spectrum of $\hat{H}_{cl}$ includes complex-valued frequencies. From (14)-(15) it follows that if $\mathbf{f}$ is an eigenfunction associated with the complex valued frequency $\omega_c = \omega' + i\omega''$ with $\omega'' > 0$ ($\mathbf{f} \leftrightarrow \omega_c = \omega' + i\omega''$) then $\mathbf{e} = \tilde{\mathbf{f}}$ is another eigenfunction associated with the complex conjugated frequency ($\mathbf{e} \leftrightarrow \omega_c = \omega' - i\omega''$) [12]. Notably, as amply discussed in Refs. [12, 13], an oscillation with frequency $\omega_c = \omega' + i\omega''$ with $\omega'' > 0$ ($\omega'' < 0$) corresponds to exponentially growing (decaying) fields, and thus to an optical instability. Therefore, this analysis demonstrates that the emergence of optical instabilities in moving media is related to a spontaneous parity-time symmetry breaking.

Crucially, when $\hat{H}_{cl}$ can be identified with a Hermitian operator its spectrum must be real-valued, and thus the $\mathcal{PT}$-symmetry is unbroken. This is guaranteed to occur in case the material matrix $\mathbf{M}$ is positive definite. In that case, we can introduce the following weighted inner product [12, 46],

$$\langle \mathbf{F}_2 | \mathbf{F}_1 \rangle = \frac{1}{2} \int d^3\mathbf{r}\, \mathbf{F}_2^* \cdot \mathbf{M}(z) \cdot \mathbf{F}_1. \tag{16}$$

It can be readily checked that because $\mathbf{M}$ is symmetric and real-valued $\langle \mathbf{F}_2 | \hat{H}_{cl} \mathbf{F}_1 \rangle = \langle \hat{H}_{cl} \mathbf{F}_2 | \mathbf{F}_1 \rangle$. Thus, in these conditions $\hat{H}_{cl}$ is Hermitian and the system has an unbroken $\mathcal{PT}$-symmetry[†]. Using (1)-(2) it can be shown that provided all the material bodies move with a speed lower than the corresponding Cherenkov emission threshold in the considered reference frame, i.e. provided $|v(z)| < c/n(z)$, the material matrix is

---

[†] Alternatively, we could define $\hat{H}_{cl} = \mathbf{M}^{-1/2} \cdot \hat{N} \cdot \mathbf{M}^{-1/2}$ so that $\hat{H}_{cl} \cdot \mathbf{U} = i\partial_t \mathbf{U}$ with $\mathbf{U} = \mathbf{M}^{1/2} \cdot \mathbf{F}$. This $\hat{H}_{cl}$ is Hermitian with respect to the canonical inner product provided the real-symmetric material matrix $\mathbf{M}$ is positive definite.

positive definite [12, 46]. Note that $\langle \mathbf{F} | \mathbf{F} \rangle = \int d^3\mathbf{r}\, W_{EM,P}$ is the total wave energy stored in the system.

Most dramatically, when the speed of at least one of the material bodies exceeds the associated Cherenkov emission threshold ($|v_i| > c/n_i$) the material matrix $\mathbf{M}$ becomes indefinite. In such a case (16) defines an *indefinite* inner product, and the property $\langle \mathbf{F}_2 | \hat{H}_{cl} \mathbf{F}_1 \rangle = \langle \hat{H}_{cl} \mathbf{F}_2 | \mathbf{F}_1 \rangle$ does not guarantee the reality of the spectrum of $\hat{H}_{cl}$, and optical instabilities may occur [12]. Therefore, the spontaneous symmetry breaking of the system requires that the speed of one or more material bodies exceeds the Cherenkov emission threshold.

## IV. Spontaneous symmetry breaking

For a system of weakly interacting dielectric bodies, the modes associated with electromagnetic instabilities and $\mathcal{PT}$-symmetry breaking can be understood as the result of the hybridization of specific guided modes supported by the individual slabs [13]. The selection rules for the interacting guided modes (*i*=1,2) impose *(i)* the matching between the frequencies ($\omega_1 = \omega_2$) and the wave vectors ($k_{x,1} = k_{x,2}$, $k_{y,1} = k_{y,2}$) of the modes, and that *(ii)* $\tilde{\omega}_1 \tilde{\omega}_2 < 0$ where $\tilde{\omega}_i$ is the oscillation frequency of the mode associated with the *i*-th slab measured in the respective co-moving frame [13]. By generalizing the results of Ref. [13] to the relativistic regime, it can be shown that the condition for spontaneous $\mathcal{PT}$-symmetry breaking is

$$|v_\Delta| > v_{\Delta,th}, \quad \text{where} \quad v_\Delta = \frac{v_2 - v_1}{1 - v_1 v_2 \frac{1}{c^2}} \tag{17}$$

represents the relative velocity of the slab 2 with respect to slab 1 in the frame co-moving with slab 1, and the threshold velocity is:

$$v_{\Delta,th} = c\frac{\frac{1}{n_1}+\frac{1}{n_2}}{1+\frac{1}{n_1}\frac{1}{n_2}}. \qquad (18)$$

This result is also consistent with a calculation of Pendry for the case of identical dielectric slabs [26]. Based on this, one can distinguish three situations. The first one is when $|v_i|<c/n_i$, $i$=1,2. In this case $\hat{H}_{cl}$ is a Hermitian operator and thus its frequency spectrum is real-valued and the wave energy is nonnegative. The second possibility is that $|v_\Delta|<v_{\Delta,th}$ and either $|v_1|>c/n_1$ or $|v_2|>c/n_2$. In this case the Hermitian symmetry is lost and the wave energy can be negative. However, the operator $\hat{H}_{cl}$ has an unbroken $\mathcal{PT}$-symmetry with a real-valued frequency spectrum. Finally, the third case occurs when $|v_\Delta|>v_{\Delta,th}$, and corresponds to a spontaneous $\mathcal{PT}$-symmetry breaking wherein the operator $\hat{H}_{cl}$ has a complex-valued frequency spectrum.

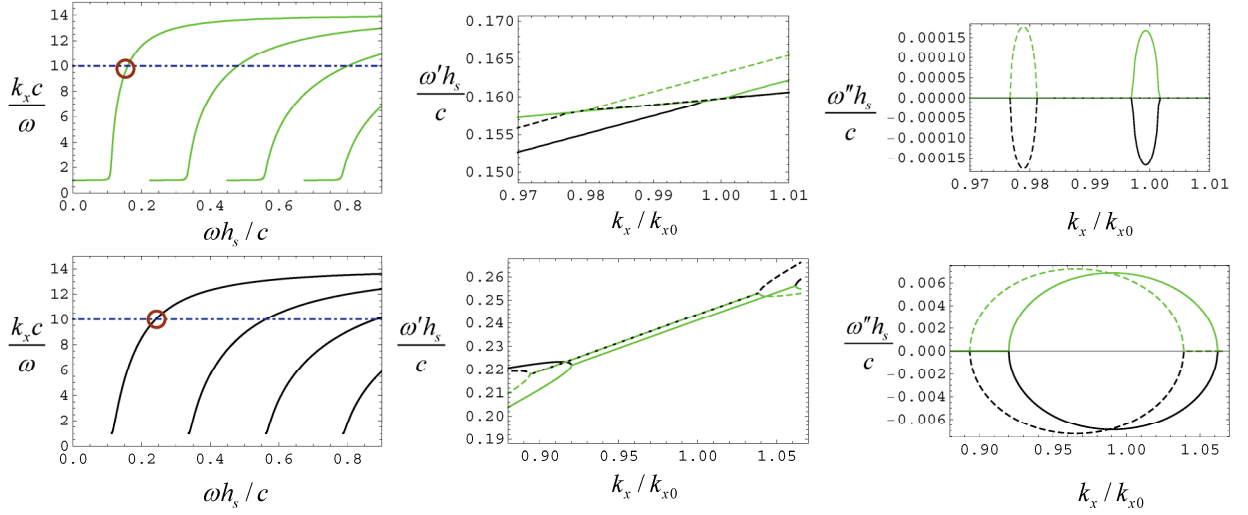

Fig. 2. (Color online) **Top row**: dispersion diagrams for *p*-polarized waves. **Bottom row**: dispersion diagrams for *s*-polarized waves. **1st column:** dispersion diagram of the guided modes for a single material slab calculated in the co-moving frame. **2nd and 3rd columns**: Dispersion $\omega' + i\omega''$ vs. $k_x$ in the lab frame for $k_y = 0$. In these plots the solid lines represent the non-relativistic calculation and the dashed lines the exact relativistic calculation. The distance between the slabs is $d = 0.75 h_s$ and $v_1 = 0$, $v_2 = c/5$ and $n_1 = n_2 = 14$. The normalization factor for the transverse wave vector is $k_{x0} = 1.60/h_s$ for *p*-polarized waves, and $k_{x0} = 2.41/h_s$ for *s*-polarized waves.

To study the properties of the guided modes for a broken $\mathcal{PT}$-symmetry, we calculated the modal dispersions $\omega = \omega' + i\omega''$ vs. $k_x$ with $k_y = 0$ for two identical finite thickness dielectric slabs with $n_1 = n_2 = 14$ and $h_s \equiv h_1 = h_2$. The slabs are backed by perfectly conducting (PEC) metallic plates and are separated by a vacuum region with thickness $d$ (see Fig. 1; the PEC plates are not shown). It is assumed that $v_1 = 0$, $v_2 = c/5$ so that the relative velocity of the two slabs is larger than the threshold for $\mathcal{PT}$-symmetry breaking. The guided modes are calculated by solving a dispersion equation written in terms of the reflection coefficients that characterize the wave scattering by the moving slabs (see

Appendix A). Generally, in the relativistic calculation the modes only split into *s* and *p* type polarizations when $k_y = 0$.

The calculated modal dispersions are shown in Fig. 2. The first column of Fig. 2 represents the dispersion $k_x$ vs. $\omega$ of the individual slabs, calculated in the pertinent co-moving frame. In our example, the instabilities result from the hybridization of the individual modes that satisfy $\omega/(k_x c) \approx v_\Delta/(2c) = 1/10$ [13]. The mode that leads to the strongest interaction is marked in the figure with a circle. The corresponding hybridized modal diagrams are shown in the 2$^{nd}$ and 3$^{rd}$ columns. As seen, because of the $\mathcal{PT}$-symmetry the complex-valued solutions occur in pairs, such that if $\omega$ is an oscillation frequency associated with an instability then $\omega^*$ also is. Figure 2 shows the dispersions calculated with both nonrelativistic (solid lines) and relativistic (dashed lines) approaches. The calculation details are described in Appendix A. Apart from a small shift, the relativistic calculation is consistent with the nonrelativistic one.

To further understand the instability properties, we calculated the field profiles for the *p*- and *s*-polarized modes for the same geometry as in Fig. 2 (see Figs. 3 and 4). The calculation is fully relativistic, and all the fields are calculated in the laboratory frame. To do this, first the fields in the vacuum gap are found, and then we use the continuity of the tangential components of the electromagnetic field at the interfaces and the relativistic field transformations [44, 48] to compute the fields in the frame co-moving with each slab. Finally, these fields are relativistically transformed back into the laboratory reference frame. The details are omitted for conciseness. In our calculations, except if stated otherwise, the fields are normalized so that $\langle \tilde{\mathbf{F}} | \mathbf{F} \rangle = \varepsilon_0 A_0 h_s \times [1 V^2/m^2]$, where $A_0 = L_x \times L_y$ represents the cross-sectional area of the slabs. The field profiles are calculated at $t = 0$. Note that because of the instability the field amplitudes vary with time.

Figures 3 and 4 show that the electromagnetic fields in the two slabs are nearly in quadrature. Interestingly, for *p*-polarization the magnetic field $H_y$ is quite small in the vacuum gap, which contrasts with the *s*-polarization case wherein the electric field $E_y$ has a significant amplitude in the gap. As already discussed in Ref. [13], this can be explained by the fact that an interface between a high-dielectric constant material and the vacuum may be seen a perfect magnetic conductor (PMC) by a *p*-polarized wave in the dielectric, and hence the *p*-polarized guided modes of the individual slabs hybridize weakly. As a consequence of this, $\omega'' = \text{Im}\{\omega\}$ is almost two orders of magnitude larger for *s*-polarized waves as compared to *p*-polarized waves.

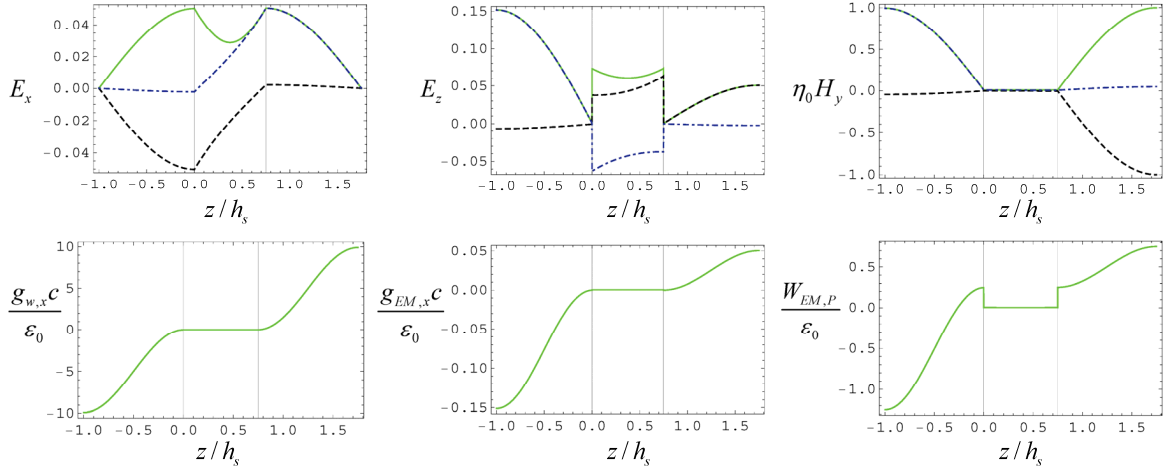

Fig. 3. (Color online) **Top row** (unities $[V/m]$): electromagnetic fields distribution for a *p*-polarized guided mode associated with a system instability ($\omega h_s / c = 0.1580 + i0.00018$ and $k_x h_s = 1.56$) and for the same configuration as in Fig. 2. The field amplitudes are normalized such that $\langle \tilde{\mathbf{F}} | \mathbf{F} \rangle = \varepsilon_0 A_0 h_s \times [1 V^2/m^2]$. Green solid lines: absolute value; Black dashed lines: real part; Blue dot-dashed lines: imaginary part. **Bottom row** (unities $[V^2/m^2]$): profiles of the normalized wave momentum density, electromagnetic momentum density, and wave energy density.

The field components normal to the interface, $E_z$ and $H_z$, are generally discontinuous. Note that even though the slabs do not have a magnetic response in the respective co-moving frame ($\mathbf{B} = \mu_0 \mathbf{H}$), in another frame there is a magnetic response and magnetoelectric coupling [Eq. (1)], and hence $\mathbf{B} \neq \mu_0 \mathbf{H}$ and $H_z$ is allowed to be discontinuous at the boundary with the moving slab ($z = 0$).

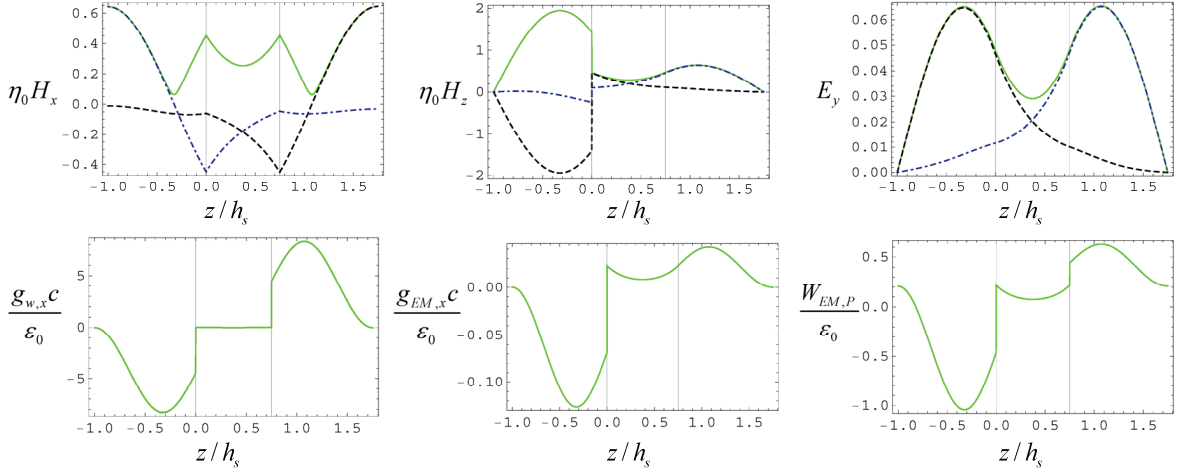

Fig. 4. (Color online) **Top row** (unities $[V/m]$): electromagnetic fields profile for a *s*-polarized guided mode associated with a system instability ($\omega h_s / c = 0.235 + 0.007i$ and $k_x h_s = 2.33$) and for the same configuration as in Fig. 2. The field amplitudes are normalized such that $\langle \tilde{\mathbf{F}} | \mathbf{F} \rangle = \varepsilon_0 A_0 h_s \times [1 V^2 / m^2]$. Green solid lines: absolute value; Black dashed lines: real part; Blue dot-dashed lines: imaginary part. **Bottom row** (unities $[V^2/m^2]$): profiles of the normalized wave momentum density, electromagnetic momentum density, and wave energy density.

Figures 3 and 4 also depict the wave energy density ($W_{EM,P}$), the *x*-component of the wave momentum density ($g_{w,x}$), and the *x*-component of the electromagnetic momentum density ($g_{EM,x}$). By numerically integrating these profiles, we verified that the total wave momentum and the total wave energy identically vanish:

$$p_{wv,x} = A_0 \int_{-h_s}^{d+h_s} g_{wv,x} dz = 0, \qquad E_{wv} = A_0 \int_{-h_s}^{d+h_s} W_{EM,P} dz = 0, \qquad (19)$$

It was analytically demonstrated in Ref. [12] that the result $E_{wv} = 0$ holds exactly for any electromagnetic mode associated with a complex-valued frequency. Similarly, it can be proven that $p_{wv,x} = 0$ always holds in the same circumstances. On the other hand, an inspection of the plots in Figs. 3 and 4 reveals that in general the total electromagnetic momentum $p_{EM,x}$ does not vanish, and hence varies with time. The rate of change of the total momentum of the system when the translational velocity of the bodies is enforced to be constant is thus $dp_{kin,x}/dt + dp_{EM,x}/dt = dp_{EM,x}/dt \neq 0$.

Significantly, $g_{EM,x}$ is two orders of magnitude smaller than $g_{wv,x}$. This demonstrates that $p_{w,i} \approx p_{ps,i}$ where $p_{ps,i} = p_{w,i} - p_{EM,i}$. Thus, the stress acting on the $i$-th slab due to the wave flow satisfies $F_i^{mat} = dp_{ps,i}/dt \approx dp_{wv,i}/dt$, consistent with the discussion in Ref. [13]. In order to enforce the velocities to be time independent, an external mechanical force is required to pump the instantaneous power $P_{ext} = -v_2 dp_{ps,2}/dt = -v_2 2\omega'' p_{ps,2} \approx -v_2 2\omega'' p_{w,2}$ into the system. It is evident from Figs. 3 and 4 that the wave momentum stored in the moving slab is negative, $p_{w,2} < 0$, and hence $P_{ext} > 0$ as it should be to pump the exponentially growing oscillations.

It is relevant to highlight that the wave energy density in the moving slab (slab 2) can be negative. This is a consequence of the requirement that to have a spontaneous $\mathcal{PT}$-symmetry breaking it is necessary that the total wave energy vanishes $E_{wv} = 0$, and hence the wave energy in slab 2 ($E_{wv,2} = \int_{slab\,2} W_{EM,P} d^3\mathbf{r}$) must be negative. The quantity $-dE_{wv,2}/dt$ has a simple physical interpretation: it is exactly the instantaneous power flowing through the interfaces of the second slab in form of electromagnetic energy, i.e. the flux of the Poynting vector through the wall $z=0$. For an analytical proof of this result see Eq. (A8) of Ref. [13]. Thus, a moving slab with negative stored energy effectively behaves as a source that pumps the rest of the system.

As discussed in Ref. [12, 13], notwithstanding $E_{wv,2} < 0$, the total energy in slab 2 is positive and grows in time. Indeed, we have [12]:

$$\frac{dE_{tot,2}}{dt} = -v_2 \frac{dp_{ps,2}}{dt} + \frac{dE_{wv,2}}{dt}. \tag{20}$$

In our example, we can write $dE_{tot,2}/dt = 2\omega''\left(-v_2 p_{ps,2} + E_{w,2}\right)$. We numerically verified that for the *s*-polarized mode represented in Fig. 4 one has $-v_2 p_{ps,2} = 0.99$ and $E_{w,2} = -0.55$ in unities of $\varepsilon_0 A_0 h_s \times \left[1V^2/m^2\right]$, confirming that $dE_{tot,2}/dt > 0$.

Notably, from the plots of $g_{EM,x}$ it is seen that the electromagnetic momentum (and hence also the electromagnetic energy) flows in opposite directions in the two slabs. In particular, the wave attached to slab 2 is a backward wave ($g_{EM,x}$ and $k_x$ have opposite signs), whereas the wave attached to slab 1 is a forward wave ($g_{EM,x}$ and $k_x$ have the same sign). Hence, the energy radiated through the interface of slab 2 to the vacuum region is dragged by the moving slab (co-propagating wave), whereas the energy radiated to the interior of slab 2 propagates in the opposite direction (counter-propagating wave). Thus, the emitted light propagates in different directions in interior and exterior of the moving slab, and this explains that the total wave momentum is conserved.

In the limit of a weak interaction, it is possible to estimate that the wave momentum stored in the *i*-th slab satisfies [12, 13] at *t=0*:

$$\left|p_{wv,1}\right| \approx \left|p_{wv,2}\right| \approx \left|\frac{k_x}{2\omega}\langle \tilde{\mathbf{F}} | \mathbf{F} \rangle\right|. \tag{21}$$

To confirm the validity of this approximation, we numerically computed the wave momentum stored in the two slabs for different velocities $v_1$ of slab 1, assuming that the relative velocity ($v_\Delta$) of the two slabs is kept constant. Thus, for each $v_1$ we solved the relativistic dispersion equation (see Appendix A) and found the values of $(\omega, k_x)$ associated with peak value of $\omega''$ for *p*-polarized waves. Note that because of the Doppler effect $(\omega, k_x)$ can vary substantially with $v_1$. In Fig. 5a we depict the numerically calculated $p_{wv,i}$ as a function of $v_1$, superimposed on the result predicted by Eq. (21). As seen, the agreement is very good. Note that as $v_1 \to -c/10$ (i.e. $v_1/|v_\Delta|+1 \to 0.5$) the wave momentum becomes quite large with the considered field normalization because $\omega' \to 0$. Even though for clarity the plot is truncated, the agreement between Eq. (21) and the numerical results is similarly good in the limit $v_1 \to -c/10$ (not shown).

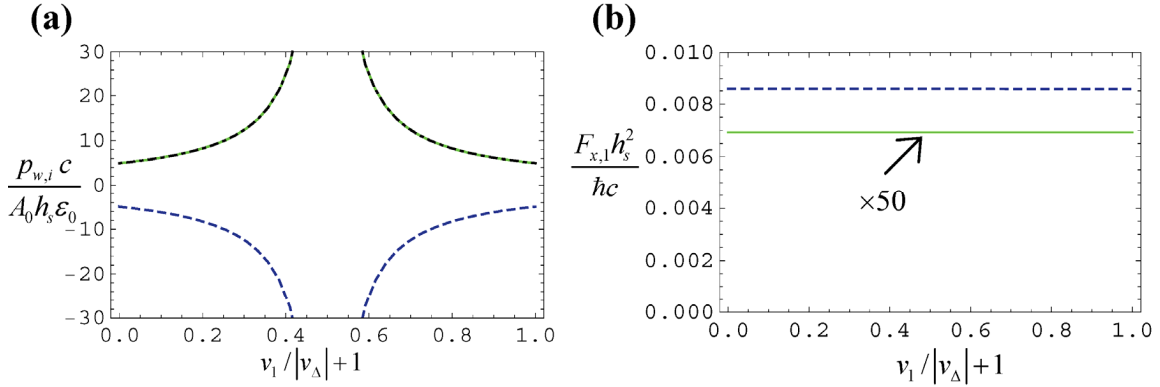

Fig. 5. (Color online) **(a)** Normalized wave momentum (in unities $[V^2/m^2]$) as a function of the normalized velocity of slab-1, $v_1$, in the lab frame. The velocity of the slab-2 in the frame of the slab-1 is kept invariant and equal to $v_\Delta = c/5$. The structural parameters are as in Fig. 2 and the fields are normalized as $\langle \tilde{\mathbf{F}} | \mathbf{F} \rangle = \varepsilon_0 A_0 h_s \times [1V^2/m^2]$. The wave is *p*-polarized. Green solid line: wave momentum in slab-1; Blue dashed line: wave momentum in slab-2; Black-dot-dashed line: analytical formula [Eq. (21)].

**(b)** Normalized force acting on slab-1 as a function of the normalized velocity of slab-1 in the lab frame with $v_\Delta = c/5$. The fields are normalized as $\langle \tilde{\mathbf{F}} | \mathbf{F} \rangle = \hbar |\omega|/2$. Green solid line: *p*-polarization; Blue dashed line: *s*-polarization;

Finally, we investigate how the stress associated with the wave flow (friction force) changes with $v_1$ in conditions similar to Fig. 5a. In this example, we suppose that the fields are normalized so that $\langle \tilde{\mathbf{F}} | \mathbf{F} \rangle = \hbar |\omega|/2$, with $\hbar$ the reduced Planck constant, consistent with the quantum vacuum normalization. Note that the field normalization depends on $v_1$ (or equivalently on the reference frame) because $\omega$ also does. Specifically, we numerically calculated $F_{x,1} = dp_{w,1}/dt = 2\omega'' p_{w,1}$ for both the *p*- and the *s*- polarized guided modes at *t*=0 (see Fig. 5b). As previously discussed, the friction force satisfies $F_1^{mat} = dp_{ps,1}/dt \approx dp_{pw,1}/dt$. Remarkably, the numerical results suggest that $F_{x,1}$ is independent of $v_1$ when the relative velocity $v_\Delta$ of the slabs is fixed. Moreover, consistent with Ref. [13] the friction force is much stronger for *s*-polarized waves, as compared to *p*-polarized waves. This happens because the *s*-polarized guided modes hybridize more strongly than the *p*-polarized waves. Note that $F_{x,1} > 0$ so that the friction force acts to reduce the relative velocity.

## V. Light amplification

One exciting opportunity created by the broken $\mathcal{PT}$-symmetry is the amplification of an optical pulse. To illustrate this, we consider the geometry of Fig. 6a, wherein a line current equidistant of the two slabs (placed at $z = z_0 = d/2$) radiates an optical pulse of finite duration. In the frequency domain, the line current is modeled by the current density $\mathbf{j}_e = I(\omega)\delta(x)\delta(z-z_0)\hat{\mathbf{y}}$ where $I(\omega)$ is the Fourier transform of the current pulse $I(t)$. It is straightforward to check that in the vacuum gap ($0 \leq z \leq d$) the electromagnetic fields in the frequency domain can be written as:

$$\mathbf{H} = \nabla \times (I\Phi\hat{\mathbf{y}}), \tag{22a}$$

$$\mathbf{E} = i\omega\mu_0 I\Phi\hat{\mathbf{y}}, \tag{22b}$$

where the scalar potential $\Phi = \Phi(x,z)$ must satisfy $\nabla^2\Phi + k_0^2\Phi = -\delta(x)\delta(z-z_0)$ with $k_0 = \omega/c$. The potential $\Phi = \Phi(x,z)$ can be written as:

$$\Phi = \Phi_0 + \Phi_{sc}, \tag{23}$$

where $\Phi_0$ is the free-space Green function,

$$\Phi_0 = \frac{i}{4}H_0^{(1)}(k_0\rho) = \frac{1}{2\pi}\int dk_x \frac{e^{-\gamma_0|z-z_0|}}{2\gamma_0} e^{ik_x x} \tag{24}$$

where $H_0^{(1)}$ is the Hankel function of 1st kind and order zero, $\rho = \sqrt{x^2 + (z-z_0)^2}$ and $\gamma_0 = \sqrt{k_x^2 - \omega^2\varepsilon_0\mu_0}$. The potential $\Phi_{sc}$ is created by the scattering of $\Phi_0$ at the interfaces of the moving dielectric slabs. Because our excitation is such that $\partial/\partial y = 0$ the radiated fields are *s*-polarized, and hence it is possible to write $\Phi_{sc}$ is terms of the electric field reflection coefficients $R_{s,i}$ at the two boundaries. Specifically, we have that:

$$\Phi_{sc} = \frac{1}{2\pi}\int dk_x \frac{1}{2\gamma_0}\left(A^+ e^{-\gamma_0(z-z_0)} + A^- e^{+\gamma_0(z-z_0)}\right)e^{ik_x x}, \tag{25}$$

where the constants $A^+, A^-$ are the solutions of the system (supposing that $z_0 = d/2$ and that the interfaces with slabs 1 and 2 are at $z = d$ and $z = 0$, respectively):

$$A^- e^{\gamma_0 d/2} = R_{s,1}(1+A^+)e^{-\gamma_0 d/2}, \tag{26a}$$

$$A^+ e^{\gamma_0 d/2} = R_{s,2}\left(1+A^-\right)e^{-\gamma_0 d/2}. \tag{26b}$$

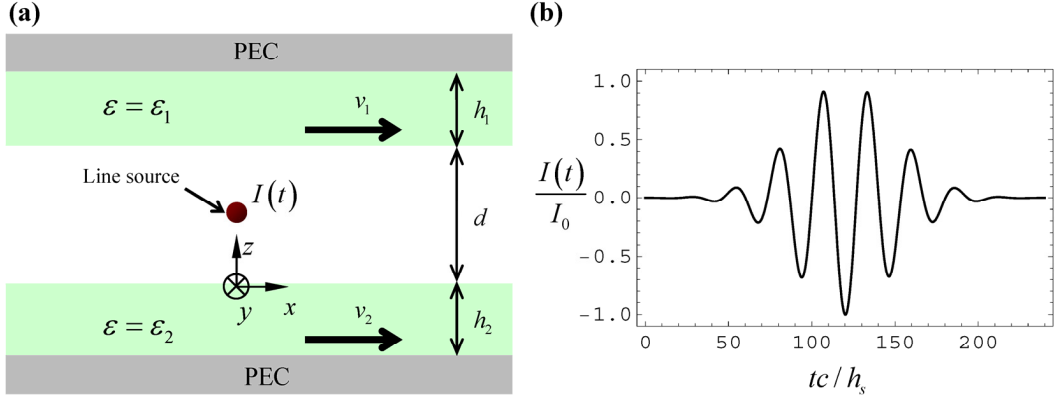

Fig. 6. (Color online) **(a)** A line source in between two moving dielectric slabs is excited by a current pulse. **(b)** The current pulse used in the numerical simulations.

The corresponding solution is:

$$A^- = \frac{R_{s,1}e^{-\gamma_0 d}\left(1+R_{s,2}e^{-\gamma_0 d}\right)}{1-R_{s,1}R_{s,2}e^{-2\gamma_0 d}}, \tag{27a}$$

$$A^+ = \frac{R_{s,2}e^{-\gamma_0 d}\left(1+R_{s,1}e^{-\gamma_0 d}\right)}{1-R_{s,1}R_{s,2}e^{-2\gamma_0 d}}. \tag{27b}$$

The reflection coefficients $R_{s,i} = R_{s,i}(\omega,k_x)$ are calculated as explained in Appendix A [Eq. (A12)]. Using Eqs. (22)-(27) and calculating the inverse Fourier transform in frequency, it is possible to write the radiated electric field as:

$$E_y(x,z,t) = \frac{1}{(2\pi)^2}\int d\omega\, e^{-i\omega t} i\omega\mu_0 I(\omega) \int dk_x \frac{1}{2\gamma_0} e^{ik_x x}\left\{e^{-\gamma_0|z-z_0|} + \frac{e^{-\gamma_0 d}}{1-R_{s,1}R_{s,2}e^{-2\gamma_0 d}}\left[R_{s,2}\left(1+R_{s,1}e^{-\gamma_0 d}\right)e^{-\gamma_0(z-z_0)} + R_{s,1}\left(1+R_{s,2}e^{-\gamma_0 d}\right)e^{+\gamma_0(z-z_0)}\right]\right\}. \tag{28}$$

The integration in $k_x$ is over the real axis. Because the system response is required to be *causal*, the inverse Fourier (Laplace) transform in frequency must be calculated over a line parallel to the real frequency axis, $\text{Im}\{\omega\} = \omega''_{int}$ such that $\omega''_{int}$ larger than the peak value of $\text{Im}\{\omega\}$ for all natural frequencies of oscillation. In other words, the system response is required to be analytic (with no poles) in the semi-plane $\text{Im}\{\omega\} > \omega''_{int}$. For the system parameters of Fig. 2 the peak value of $\text{Im}\{\omega\}$ is estimated to be $\max \omega'' h_s / c = 0.007$ (mode represented in Fig. 4). In our calculations the inverse transform was evaluated along the line defined by $\omega''_{int} h_s / c = 0.012$.

The current excitation pulse in the time domain is taken to be of the form:

$$I(t) = I_0 \text{Re}\left\{ e^{-(t-t_d)^2/(2\sigma_I^2)} e^{-i\omega_0 t} \right\}, \tag{29}$$

where $I_0$ is the peak current, $\omega_0$ is the frequency of oscillation, $t_d$ is roughly the instant wherein the current pulse is peaked, and $\sigma_I$ determines the duration of the pulse. In the numerical simulations the parameters used were $\sigma_I c / h_s = 30$, $t_d = 4\sigma_I$ and $\omega_0 h_s / c = 0.235$. Note the value of $\omega_0$ matches the value of the real part of the oscillation frequency associated with the system instability in Fig. 4. The profile of the current pulse in the time domain is represented in Fig. 6b. In the spectral domain the current is given by $I(\omega) = \int dt\, I(t) e^{i\omega t}$, which can be written as:

$$I(\omega) = I_0 \sigma_I \sqrt{2\pi} \frac{1}{2} \sum_{\pm} e^{i(\omega \pm \omega_0) t_d} e^{-\frac{\sigma_I^2 (\omega \pm \omega_0)^2}{2}}. \tag{30}$$

Using this formalism we computed the emitted electric field $E_y$ at the interface with the moving slab ($z = 0$) for a system with the same parameters as in Fig. 2 and for $v_1 = 0$. The profiles of $E_y$ as a function of time for different positions ($x = const.$) along the structure are represented in Fig. 7 for two cases *(i)* $v_2 = c/5$ (first row in Fig. 7) and *(ii)* $v_2 = 0$ (second row in Fig. 7).

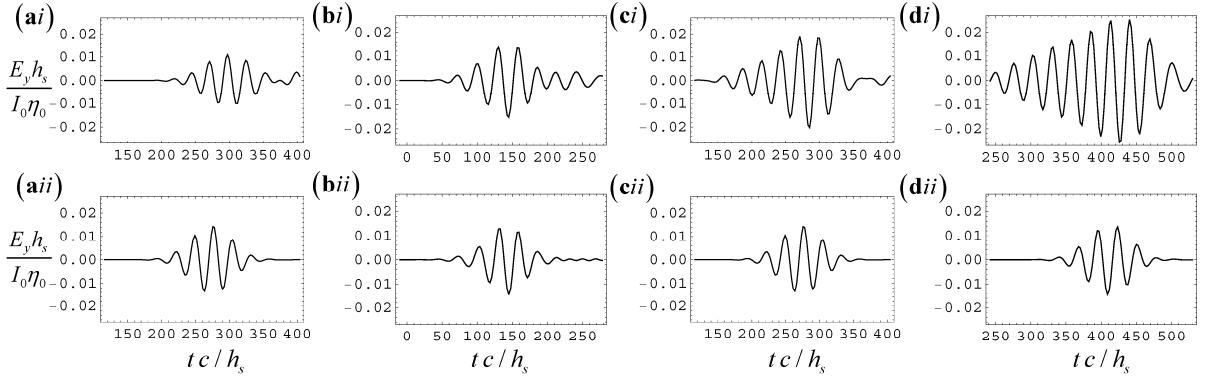

Fig. 7. (Color online) Time evolution of the field ($E_y$) radiated by a current pulse with finite duration placed at the origin. The distance between the slabs is $d = 0.75h_s$, $v_1 = 0$, and $n_1 = n_2 = 14$. The field $E_y$ is calculated at the interface with the second slab and for **(a)** $x = -10h_s$, **(b)** $x = h_s$, **(c)** $x = 10h_s$, and **(d)** $x = 19h_s$. The results of the first row are for the case (i) $v_2 = c/5$ whereas the results of the second row are for (ii) $v_2 = 0$. The development of a field instability due to the spontaneous parity-time symmetry breaking is evident in case (i).

As seen, when both slabs are at rest (i.e. $v_2 = 0$, corresponding to the second row in Fig. 7) the excited optical field is guided along the structure and the peak amplitude is roughly the same for all observation points. The propagation velocity of the optical pulse is of the order $c/n$ indicating that most of the energy propagates in the dielectric slabs. Rather different, when the relative velocity of the two slabs exceeds the threshold for a spontaneous $\mathcal{PT}$-symmetry breaking (i.e. $v_2 = c/5$, corresponding to the first row in Fig. 7), the optical pulse is amplified as it propagates in the waveguide. Indeed, comparing Fig. 7(bi) with Fig. 7(biv) it is seen that after a propagation distance of only $18h_s$ the amplitude of the pulse is increased by a factor of roughly 1.7. Moreover, comparing the first and second rows of Fig. 7, it is evident that the pulse arrives earlier at observation points with $x > 0$ when $v_2 = c/5$. This is explained by the fact that $v_2$ is larger than $c/n$ by a factor of 2.8 and hence the moving slab drags the emitted optical field so that the velocity of propagation is increased. These results suggest the possibility of having an optical amplifier based on the relative motion of two polarizable bodies.

## VI. Spontaneous $\mathcal{PT}$-symmetry breaking in dispersive moving systems

Up to now, we considered systems formed by nondispersive dielectrics. It was seen that to have a broken $\mathcal{PT}$-symmetry it is necessary that the relative velocity of the two slabs (i.e. the media wherein light propagates) exceeds the speed of light in the slabs. This is a quite strict requirement because for nondispersive dielectrics the speed of light is $c/n$, which in practice, for most dielectrics, is only a few times smaller than the speed of light in vacuum.

It is possible to somewhat alleviate this problem by considering dispersive media [14, 26]. In theory, dispersive media may allow for wave propagation with arbitrarily low phase velocities, and thus the velocity threshold for a broken $\mathcal{PT}$-symmetry can be much less than the speed of light, so that the wave instabilities may be possibly observed with non-relativistic velocities.

The response of a dispersive dielectric is characterized in the co-moving frame by a frequency dependent permittivity $\varepsilon = \varepsilon(\tilde{\omega})$ where $\tilde{\omega}$ is the frequency measured in the co-moving frame (for simplicity in what follows it is assumed that $\mu = \mu_0$). Similar to what was already discussed in Sect. II, the same material is seen in a different inertial reference frame as a bianisotropic medium. The constitutive relations (1)-(2) remain valid in the frequency domain. However, now the frequency $\tilde{\omega}$ must be written in terms of the frequency $\omega$ in the laboratory frame using the relativistic Doppler shift formula [48]:

$$\tilde{\omega} = \gamma_v \left( \omega - v k_x \right), \qquad \text{with} \quad \gamma_v = 1/\sqrt{1 - (v/c)^2} \;. \tag{31}$$

Therefore, the material matrix in the laboratory frame depends both on the frequency ($\omega$) and on the wave vector ($k_x$), so that the response of a moving material is characterized by both frequency and spatial dispersion. Hence, because we are interested in z-stratified structures, in our problem $\mathbf{M} = \mathbf{M}(z; \omega, k_x)$.

For frequency dispersive systems the time evolution problem cannot be reduced to a Schrödinger-type equation, and hence the analogy with quantum theory is imperfect. However, as shown next, the main ideas of Sect. III are recovered in the spectral domain. Indeed, using the reality condition $\varepsilon^*(\tilde{\omega}) = \varepsilon(-\tilde{\omega}^*)$, it is straightforward to verify that:

$$\mathbf{M}(z; -\omega^*, -k_x^*) \cdot \mathcal{PT} = \mathcal{PT} \cdot \mathbf{M}(z; \omega, k_x). \tag{32}$$

To make further progress it is necessary to assume that the materials are lossless so that $\varepsilon(\tilde{\omega}) = \varepsilon(-\tilde{\omega})$. In that case, we can write:

$$\mathbf{M}(z; \omega^*, k_x^*) \cdot \mathcal{PT} = \mathcal{PT} \cdot \mathbf{M}(z; \omega, k_x), \quad \text{(lossless system)}. \tag{33}$$

Hence, using Eq. (12a) it is seen that in the spectral domain $\hat{H}_{cl}(\omega, k_x) = \mathbf{M}^{-1} \cdot \hat{N}$ satisfies:

$$\hat{H}_{cl}(\omega^*, k_x^*) \cdot \mathcal{PT} - \mathcal{PT} \cdot \hat{H}_{cl}(\omega, k_x) = 0. \tag{34}$$

This result is the counterpart of $\left[\hat{H}_{cl}, \mathcal{PT}\right]=0$ for nondispersive systems. From here, we see that if $\hat{H}_{cl}(\omega, k_x) \cdot \mathbf{F}_{\omega,k_x} = \omega \mathbf{F}_{\omega,k_x}$ then $\hat{H}_{cl}(\omega^*, k_x^*) \cdot \mathcal{PT} \cdot \mathbf{F}_{\omega,k_x} = \omega^* \mathcal{PT} \cdot \mathbf{F}_{\omega,k_x}$. This property is the analogue of Eqs. (14)-(15) and establishes that the frequency spectrum of a system formed by lossless dispersive moving slabs has a mirror symmetry with respect to the real-frequency axis, i.e. the complex eigenfrequencies occur in complex conjugated pairs and the corresponding eigenvectors are related by the $\mathcal{PT}$-transformation.

In order to numerically confirm these results, we computed the dispersion diagrams of the natural modes supported by two identical metallic half-spaces in relative motion ($h_1 = h_2 \to \infty$ in Fig. 1). The material response is modeled by a Drude-dispersion model with $\varepsilon_m / \varepsilon_0 = 1 - 2\omega_{sp}^2 / \left[\omega(\omega+i\Gamma)\right]$, where $\Gamma > 0$ is the collision frequency and $\omega_{sp}$ is the surface plasmon resonance frequency such that $\text{Re}\{\varepsilon_m(\omega_{sp})/\varepsilon_0\} = -1$. In the simulations we used $\omega_{sp}/2\pi = 646\,\text{THz}$ which is expected to model the response of silver [49].

In the first example, we computed the modal dispersion $\omega = \omega' + i\omega''$ vs. $k_x$ with $k_y = 0$ for a lossless system $\Gamma = 0^+$ when the vacuum gap thickness is $d = 10\,nm$, $v_1 = 0$, and $v_2 = 2\omega_{sp}d = 0.27c$. For the details of the modal diagram calculations see Ref. [14]. As shown in Figs. 8a and 8b, the frequency spectrum is complex valued. Moreover, the complex poles occur in complex conjugated pairs, which is the signature of a broken $\mathcal{PT}$-symmetry. Similar to Fig. 2, the relativistic calculation (dashed lines) is consistent with the non-relativistic one (solid lines). Notably, the instability is more broadband in presence of material dispersion, and even more interesting, the peak value of $\omega''/\omega'$ is several orders of magnitude larger in this example. This indicates that the presence of material dispersion promotes, indeed, the enhancement of the system instabilities.

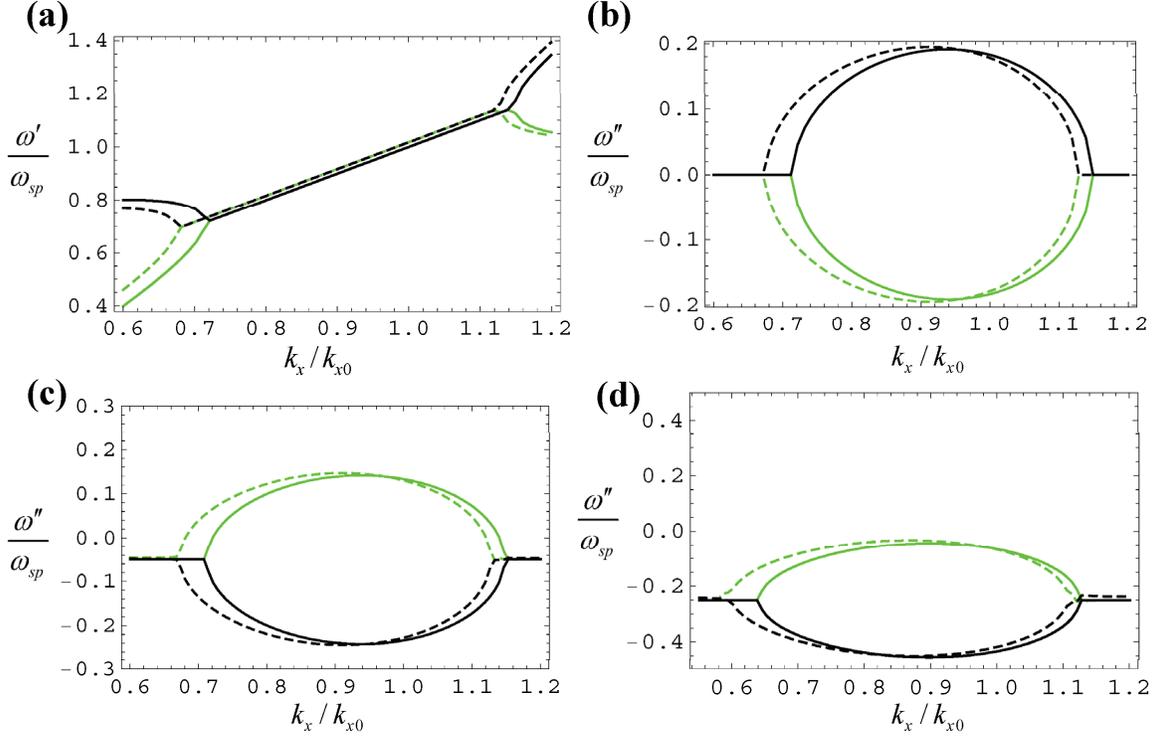

Fig. 8. (Color online) Two silver semi-spaces ($h_1 = h_2 \to \infty$) separated by a vacuum gap with thickness $d = 10nm$ [$\omega_{sp}d/c = 0.135$] are in relative motion. **(a)** Real part and **(b)** imaginary part of oscillation frequencies of two natural modes (black and green lines) as a function of the normalized $k_x$ for $\Gamma = 0$, and $v_2 = 2\omega_{sp}d$. The transverse wave number is normalized to $k_{x0} = 2\omega_{sp}/v_2$. In all the plots, $k_y = 0$, and the solid lines represent the non-relativistic calculation and the dashed lines the exact relativistic calculation. **(c)** and **(d)**: Similar to **(b)** but for $\Gamma = 0.1\omega_{sp}$ and $\Gamma = 0.5\omega_{sp}$, respectively.

In presence of material loss, Eq. (34) is no longer valid, and hence the complex valued natural oscillation frequencies do not need to occur in complex-conjugated pairs. This property is illustrated in Figs. 8c and 8d for the cases $\Gamma = 0.1\omega_{sp}$ and $\Gamma = 0.5\omega_{sp}$, respectively (see also Ref. [14]). Note that the system can support exponentially growing oscillations with $\omega'' > 0$ even in presence of realistic material loss (see Fig. 8c; for silver $\Gamma \approx 0.2\omega_{sp}$, see Ref. [14]).

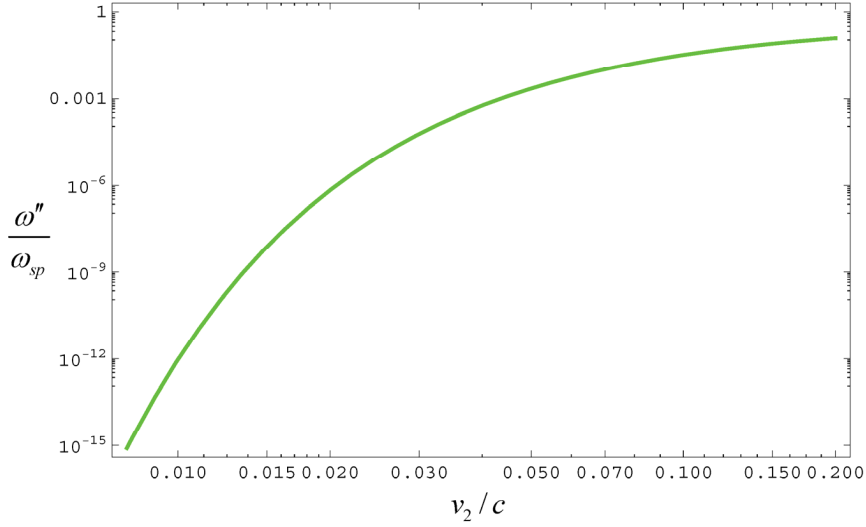

Fig. 9. (Color online) Similar to Fig. 8b, but $\omega''$ is calculated as a function of $v_2$ for $k_x = 2\omega_{sp}/v_2$, $\Gamma = 0$, and $k_y = 0$. The calculation is non-relativistic and only the mode with $\omega'' > 0$ is represented.

Crucially, for $\omega = \omega_{sp}$ the two metallic semi-spaces support surface plasmons with arbitrarily small phase velocities $v_{ph} = \omega/k_x \to 0$ when $\omega \to \omega_{sp}$ and $\Gamma = 0^+$. Hence, in the continuous lossless limit, the threshold velocity for a spontaneous parity-time symmetry breaking vanishes: $v_{\Delta,th} = 0$. This property is illustrated in Fig. 9. However, in presence of material loss the threshold velocity for a system instability is evidently nonzero because the phase velocity has a lower bound and because the "gain" must surpass the absorption. As shown in Ref. [14], for the realistic silver loss and $d = 10nm$ the threshold velocity is as large as $v_{\Delta,th} \sim \omega_{sp}d = 0.135c$.

-31-

# VII. Conclusion

It was demonstrated when the relative velocity of two sheared lossless dielectrics exceeds a certain threshold, the time evolution of the electromagnetic waves is described by a non-Hermitian parity-time symmetric operator. For sufficiently large relative velocities, the system may enter into a broken parity-time symmetry regime and electromagnetic instabilities may arise due to the spontaneous conversion of kinetic energy into light. We characterized the properties of the electromagnetic fields associated with a exponentially growing wave instability. It was shown that in the reference frame wherein one of the slabs is at rest, the light emitted towards the exterior of the moving slab co-propagates with the moving slab, whereas the light emitted towards the interior of the moving slab propagates in the opposite direction, such that the total wave momentum is preserved. We studied the time evolution of the electromagnetic field in presence of a light source in the broken $\mathcal{PT}$ - regime, and showed that the emitted optical pulse is amplified as it propagates in the structure. Moreover, it was proven that for dispersive lossless dielectrics the parity-time symmetry breaking may have no velocity threshold. Finally, we would like to note that even though our study deals with light propagation in material media, similar effects are expected to occur in other physical systems (e.g. for sound waves) wherein two slabs that support wave propagation are sheared with a relative velocity exceeding twice the wave velocity in the media.

## *Appendix A: The modal equation*

In this Appendix, we derive the dispersion equations used to characterize the natural oscillations of the system in Fig. 1.

We start by considering that the field variation in the $x$ and $y$ coordinates is of the form $e^{i(k_x x + k_y y)}$. It was shown in Ref. [14] that the transverse electric fields evaluated in the frame co-moving with the pertinent medium ($\tilde{E}_x, \tilde{E}_y$) are related to the corresponding fields evaluated in the laboratory frame ($E_x, E_y$) as:

$$\begin{pmatrix} \tilde{E}_x \\ \tilde{E}_y \end{pmatrix} = (\mathbf{1} + \mathbf{A}) \cdot \begin{pmatrix} E_x \\ E_y \end{pmatrix}. \tag{A1}$$

where $\mathbf{1}$ is the identity matrix and

$$\mathbf{A}(\omega, k_x, k_y, v) = \gamma_v \begin{pmatrix} 0 & 0 \\ \beta c \dfrac{k_y}{\omega} & 1 - \dfrac{1}{\gamma_v} - \beta c \dfrac{k_x}{\omega} \end{pmatrix}, \tag{A2}$$

with $\beta = v/c$ and $\gamma_v = 1/\sqrt{1 - v^2/c^2}$. In this Appendix, the quantities with a tilde hat are calculated in the frame co-moving with the medium where the material response is described by the parameters $\varepsilon, \mu$. Because the $(c\mathbf{D}, \mathbf{H})$ fields are relativistically transformed in the same manner as the $(\mathbf{E}, c\mathbf{B})$ fields [43], the transverse magnetic fields are also linked by:

$$\begin{pmatrix} \tilde{H}_x \\ \tilde{H}_y \end{pmatrix} = (\mathbf{1} + \mathbf{A}) \cdot \begin{pmatrix} H_x \\ H_y \end{pmatrix}. \tag{A3}$$

Let us consider a plane wave propagating in the medium, and introduce characteristic impedance matrices such that in the co-moving frame we can write

$$\begin{pmatrix} \tilde{E}_x \\ \tilde{E}_y \end{pmatrix} = Z_c^{co}(\tilde{\omega}, \tilde{k}_x, \tilde{k}_y) \cdot \begin{pmatrix} \tilde{H}_x \\ \tilde{H}_y \end{pmatrix}, \tag{A4}$$

whereas in the laboratory frame we have

$$\begin{pmatrix} E_x \\ E_y \end{pmatrix} = Z_c(\omega, k_x, k_y) \cdot \begin{pmatrix} H_x \\ H_y \end{pmatrix}. \tag{A5}$$

The frequencies and transverse wave numbers in the two frames are linked by the relativistic Doppler shift formulas [48],

$$\tilde{\omega} = \gamma_v (\omega - v k_x), \qquad \tilde{k}_x = \gamma_v (k_x - \omega v / c^2), \qquad \tilde{k}_y = k_y. \tag{A6}$$

Then, it is evident that:

$$Z_c (\omega, k_x, k_y) = (1+\mathbf{A})^{-1} \cdot Z_c^{co} (\tilde{\omega}, \tilde{k}_x, \tilde{k}_y) \cdot (1+\mathbf{A}). \tag{A7}$$

It can be verified that in the co-moving frame:

$$Z_c^{co} (\tilde{\omega}, \tilde{k}_x, \tilde{k}_y) = \frac{1}{\tilde{\omega} \varepsilon \tilde{k}_z} \begin{pmatrix} \tilde{k}_x \tilde{k}_y & \tilde{k}_y^2 + \tilde{k}_z^2 \\ -\tilde{k}_x^2 - \tilde{k}_z^2 & -\tilde{k}_x \tilde{k}_y \end{pmatrix}, \tag{A8}$$

where $\tilde{k}_z = \sqrt{\tilde{k}_x^2 + \tilde{k}_y^2 - \tilde{\omega}^2 \varepsilon \mu}$ is the *z*-propagation constant for a plane wave.

For simplicity, in this article we restrict our attention to waves with $k_y = 0$ so that the normal modes can be split into *s* and *p* polarized modes. It can be checked that for $k_y = 0$ the characteristic impedance matrix is anti-diagonal, so that it is possible to introduce two scalar characteristic impedances for *s* and *p* polarized waves:

$$Z_c^s = -\frac{E_y}{H_x}, \qquad Z_c^p = \frac{E_x}{H_y}. \tag{A9}$$

Explicit calculations show that:

$$Z_c^s = \frac{\tilde{k}_x^2 + \tilde{k}_z^2}{\tilde{\omega} \varepsilon \tilde{k}_z} \frac{1}{\gamma_v \left(1 - \beta c \frac{k_x}{\omega}\right)}, \qquad Z_c^p = \frac{\tilde{k}_z}{\tilde{\omega} \varepsilon} \gamma_v \left(1 - \beta c \frac{k_x}{\omega}\right). \tag{A10}$$

Next, we note that the *z*-propagation constant of a plane wave is the same in the frame co-moving with the slab and in the laboratory frame. Hence, it follows that in the relevant medium:

$$k_z = \tilde{k}_z = \sqrt{\gamma_v^2 (k_x - \omega v / c^2)^2 + k_y^2 - \gamma_v^2 (\omega - v k_x)^2 \varepsilon \mu}. \tag{A11}$$

where we used the Doppler shift formulas (A6).

Using Eqs. (A10)-(A11) and standard transmission line theory, it is now straightforward to find the reflection coefficient for a plane wave incident on the *i*-th interface. For the case wherein the relevant slab is backed by a PEC plate and has thickness $h_i$ it is found that:

$$R_{l,i} = \frac{-iZ_{c,i}^l \tan(\tilde{k}_{z,i} h_i) - Z_{c,0}^l}{-iZ_{c,i}^l \tan(\tilde{k}_{z,i} h_i) + Z_{c,0}^l}, \qquad l=s,p \qquad (A12)$$

The quantities with subscript *i* are evaluated at the slab side of the interface, and the quantities with subscript "0" are evaluated at the vacuum side.

Following Refs. [13, 14], the dispersion equation for the natural oscillation frequencies in an air cavity with thickness *d* and delimited by two moving dielectric slabs is given by

$$1 - e^{-2\gamma_0 d} R_{l,1}(\omega, k_x, v_1) R_{l,2}(\omega, k_x, v_2) = 0, \qquad (A13)$$

where *l=s,p* determines the field polarization, $\gamma_0 = \sqrt{k_x^2 - \omega^2/c^2}$, and $R_{i,l}(\omega, k_x, v_i)$ (*i*=1,2) is the reflection coefficient for the *i*-th interface defined as in Eq. (A12). The non-relativistic calculations of Fig. 2 are based on the approximation $R_l(\omega, k_x, v) \approx R_l(\omega - k_x v, k_x, 0)$ in Eq. (A13).

**Acknowledgements:** This work is supported in part by Fundação para a Ciência e a Tecnologia grant number PTDC/EEI-TEL/2764/2012. The author acknowledges fruitful discussions with A. Alù.